\documentclass[twocolumn]{jpsj2} 
\def\runauthor{Y. H. Matsuda et al.,}

\title{%
High field x-ray diffraction study on a magnetic-field-induced valence transition in 
YbInCu$_4$
}

\author{%
Y. H. Matsuda\thanks{E-mail address: ymatsuda@cc.okayama-u.ac.jp}, T. Inami$^1$, K. Ohwada$^1$, Y. Murata, 
H. Nojiri$^2$, Y. Murakami$^{3,1}$, H. Ohta$^4$, W. Zhang$^4$, K. Yoshimura$^4$
}
\inst{%
Department of Physics, Faculty of Science, Okayama University, 3-1-1 Tsushimanaka, Okayama, 700-8530, Japan \\
$^1$Synchrotron Radiation Research Center, Japan Atomic Energy Research Institute, Mikazuki-cho, Hyogo 679-5148, Japan \\
$^2$Institute for Materials Research, Tohoku University, 2-1-1 Katahira, Aoba-ku, Sendai 980-8577, Japan \\
$^3$Graduate School of Science, Tohoku University, Sendai 980-8578, Japan \\
$^4$Department of Chemistry, Graduate School of Science, Kyoto University, Kyoto 606-8502, Japan 
}
\recdate{\today}

\abst{
We report the first high-field x-ray diffraction experiment using synchrotron x-rays and pulsed 
magnetic fields exceeding 30 T. Lattice deformation due to a magnetic-field-induced valence 
transition in YbInCu$_4$ is studied.
It has been found that the Bragg reflection profile at 32 K changes significantly at around 27 T  
due to the structural transition.
In the vicinity of the transition field the low-field and the high-field phases are observed simultaneously 
as the two distinct Bragg reflection peaks: 
This is a direct evidence of the fact that the field-induced valence state transition is the first 
order phase transition. The field-dependence of the low-field-phase Bragg peak intensity is found to be 
scaled with the magnetization.

}

\kword{x-ray diffraction, pulsed magnetic field, valence transition

}

\begin{document}
\maketitle

\section{Introduction}
A magnetic field is a key parameter to control two competing aspects of correlated electrons: 
itinerancy and localization.  This competition causes various field-induced transitions, such as 
metamagnetism in uranium compounds \cite{URu2Si2} or valence state transition in intermetallic 
substances \cite{Ce, Eu, Yb0}, which are interesting subjects in magnetism.
One of the most well known examples of the valence state transition is the $\gamma$-$\alpha$ transition in Ce.\cite{Ce} 
The $f$-electron state, localized at room temperature, becomes itinerant at low temperatures due to the strong 
hybridization. Such valence state transition is also caused by application of a high pressure or a high magnetic 
field. In fact, the itinerant $\alpha$-state is induced by high pressures. \cite{Ce}
On the other hand, the field-induced $\alpha$-$\gamma$ transition is predicted to occur at 
around 2000 T \cite{2000T}and the real experiment in such high fields is almost impossible to be performed. 
Although the mechanism of the valence state transition is not fully understood,
the field induced valence state transition is considered to be closely related to the collapse of the Kondo singlet 
and the metamagnetism is observed.

Among the Kondo materials, YbInCu$_4$ is another model substance of the valance transition 
and has been extensively studied  because the transition is very sharp. \cite{Yb0}  
The valence state changes from Yb$^{3+}$ to Yb$^{2.8+ \sim 2.9+}$ at around $T_v$=42 K.  \cite{Yb1}   
The valence change was determined 
by the change in the lattice volume and by the x-ray absorption spectra near the $L$-edge of the Yb ions. \cite{Yb3, Yb4}
Above $T_v$, the magnetic moments are nearly localized and the Currie paramagnetic behavior is observed. 
Below $T_v$, the local magnetic moments seem to vanish due to the Kondo effect and the enhanced Pauli 
paramagnetism appears with anomalies in conductivity and specific heat.
This material is also known to undergo a sharp field-induced valence state transition. \cite{Yoshimura}
The  transition occurs in experimentally accessible magnetic fields and has been studied 
by a metamagnetic transition or a sharp increase of the conductivity .\cite{Yoshimura, MR, Zhang}
An advantage to study the field induced valence transition compared to the temperature driven transition 
is that an experiment can be performed at a low temperature.
Hence, the effect of thermal fluctuations is suppressed and a clear transition is examined.

One of the most important parameters in the valence state transition is a change of the lattice volume. 
According to the literature \cite{C15b}, when temperature increases the volume decreases by 0.45\% 
at $T_v$ keeping the $C$15$b$ type cubic structure.
The magnetostriction measurement also indicates that the field-induced transition is associated with the lattice 
contraction. \cite{Yoshimura} The x-ray diffraction is a powerful method to investigate the details of the transition,  
because the change in the crystal lattice is measured directly. Moreover, we can observe the contributions of the 
two phases separately as two Bragg reflection peaks in the vicinity of the phase transition. 
Especially x-ray diffraction study on the isothermal field-induced valence transition 
is an interesting experiment to clarify the mechanism of the valence transition.
However, the high field x-ray diffraction experiment had never been made so far for experimental difficulties.

Recently, we have developed the high-magnetic-field x-ray diffraction measurement system 
using a compact pulsed field generator and synchrotron x-rays.\cite{Matsuda} 
By this system it is possible to observe the change in the crystal lattice only with the single pulse because of 
the strong intensity of the synchrotron x-rays.
The field dependence of the diffraction intensity is traced in real time and it can be related directly to 
other quantities such as magnetization measured in a pulsed magnetic field.
In this paper we report the first high field x-ray diffraction experiment above 30 T.
The lattice deformation in YbInCu$_4$ due to the field-induced valence transition is examined in detail ; 
the coexistence of the two valence state phases and microscopic nucleation in the phase transition 
are discussed.

\section{Experiments}
\begin{figure}[t]
\begin{center}
\includegraphics[width=9.5cm]{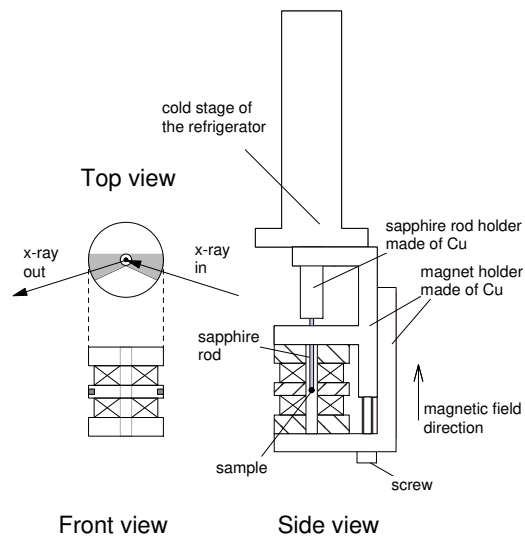}
\end{center}
\caption{Schematic view of the experimental setup around the magnet. }
\label{setup}
\end{figure}

The experiment was performed at beam line BL22XU of SPring-8.
Figure~\ref{setup} shows a schematic view of the experimental setup around the sample.
The small pulsed magnet is wound by a CuAg wire and magnetic fields up to 33 T are generated with the pulse duration 
of 0.6 ms.
The magnet is driven by a portable capacitor bank \cite{Matsuda} and the required energy for 33 T 
is about 1 kJ, 
which is only 1 \% 
of a conventional pulsed field generator.

The outer dimensions of the magnet are 20 mm in diameter and 24 mm in height.
The bore is 3 mm and a sample is glued on the sapphire rod of 1 mm in diameter. 
The rod is attached to the cold stage of a helium-gas-closed-cycle refrigerator and temperature 
as low as 10 K is achieved.
The sample temperature is monitored by a thermocouple placed on the sapphire rod near the sample.
The magnet is so small that it can be cooled by the same refrigerator.
As shown in the cross sectional top view of the magnet, there are two 25-degree-wide windows for the incident and 
the diffracted x-rays. The height of the windows is 1.6 mm and the full vertical  allowance is 9 degree.
The refrigerator is attached to the $\chi$-circle of a standard four circle diffractometer.

The photon energy of synchrotron x-rays used is 8.5 keV($\lambda$=1.457 \AA). 
A time domain photon counting technique is used to monitor the diffraction intensity 
in a pulsed magnetic field.\cite{Matsuda, Inami}
The typical time resolution is 10 $\mu$s, which is achieved by using an avalanche photo diode (APD) \cite{Kishimoto} 
and a multi channel scalar (MCS). The corresponding magnetic field resolution $\Delta B$ for $\Delta t=10 \mu s $ 
depends on the field position;  for example , $\Delta B \sim$ 0.8 T to 0.2 T at $B$=25 T to 30 T.

A single crystal of YbInCu$_4$ grown from In-Cu flux was used for the measurements. 
The sample is shaped into a small rectangle ($\sim$ $1 \times 1 \times 1$ mm$^3$).

\section{Results }

\begin{figure}[t]
\begin{center}
\includegraphics[width=7cm]{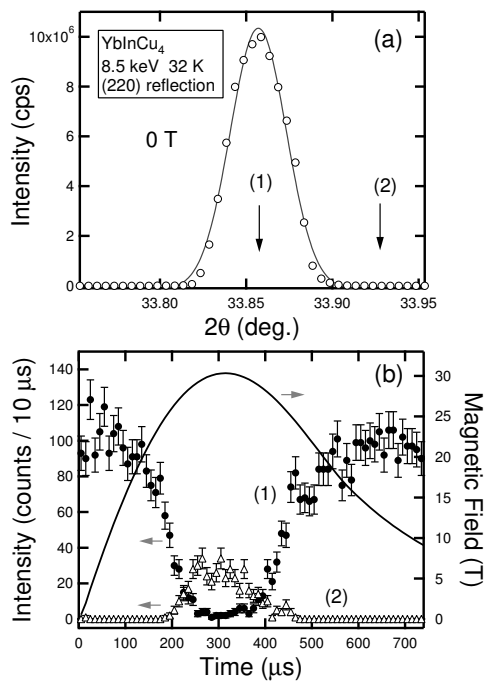}
\end{center}
\caption{
(a) The $\theta-2\theta$ (220) Bragg reflection profile of YbInCu$_4$ at 0 T. 
(b) The x-ray diffraction intensities and the magnetic field as a function of time. 
The closed circles show the (220) Bragg reflection intensity at the peak angle of the reflection 
profile at 0 T, while the open triangles show the reflection intensity at the higher angle by 
0.07$^{\circ}$.}
\label{time_domain}
\end{figure}

\begin{figure}[t]
\begin{center}
\includegraphics[width=7cm]{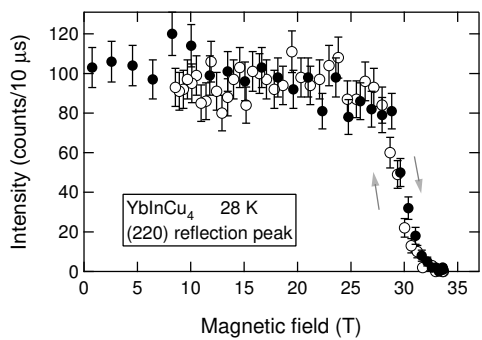}
\end{center}
\caption{The magnetic field dependence of the reflection intensity of YbInCu$_4$ at the peak 
angle at 28 K. The closed circles denote the data obtained for the up-sweep of the magnetic field, 
while the open circles are the data obtained for the down-sweep of the magnetic field.}
\label{field _dep}
\end{figure}

The $\theta-2\theta$ profile of the (220) Bragg reflection at zero magnetic field is shown in 
Fig.~\ref{time_domain} (a) .  The full width at half maximum of the peak is about $0.05^{\circ}$.
Figure~\ref{time_domain} (b) shows the time dependence of the magnetic field and that of the Bragg reflection 
intensity.
The traces (1) and (2) correspond to the data obtained at the angles (1) and (2) shown in 
Fig.~\ref{time_domain} (a), respectively.
The intensity at the angle (1) shows a pronounced decrease between 20 T $\sim$ 27 T and simultaneously the intensity 
at the angle (2) increases.
Since the metamagnetic transition occurs in the same field range \cite{Mush2} the observed changes
in the reflection intensities are considered to be the direct evidence of the lattice shrinkage due to 
the valence state transition.
From the time domain data, a field dependence of the Bragg reflection intensity is obtained 
as shown in Fig.~\ref{field _dep}.
The transition field $H_v$ is found to be around 28 T at this temperature.
Note that such field dependence can be measured using only one shot of the pulsed magnetic field for 
the strong intensity of the synchrotron x-rays.

Here we mention that the sample temperature seems not to increase significantly due to the eddy 
current by the pulsed magnetic field. We confirmed that $H_v$ is nearly unchanged when we change the sweep rate 
of the magnetic field. The estimated temperature rise of the sample during the pulsed field 
is at most around 1 K for the measurement temperature of 30 K. 
Making the cross section of the sample small is important to suppress 
the eddy current heating.

\begin{figure}[t]
\begin{center}
\includegraphics[width=7cm]{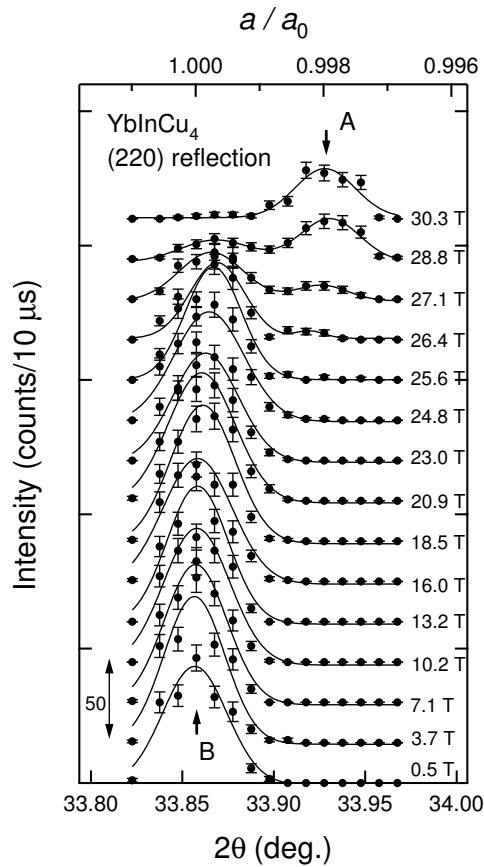}
\end{center}
\caption{The (220) reflection peak profiles at 32 K and various magnetic fields. The solid curves 
are the results of a fitting analysis using the Gaussian curve. The data obtained in the up-sweep of the pulsed field 
are plotted. }
\label{profile}
\end{figure}

To trace the change in the Bragg reflection peak profile, the measurements were repeated at 
different diffraction angles. 
The Bragg reflection profiles at different magnetic fields are obtained by re-plotting the set of data as a function of the diffraction angle as shown in Fig.~\ref{profile}:
It clearly shows the magnetic field dependence of the (220) reflection profile at 32 K.
The vertical line of the data points corresponds to the reflection intensity variation 
as shown in Fig.~\ref{field _dep}. 

In Fig.~\ref{profile} we found that a new reflection peak (denoted as peak A) appears at larger diffraction angle 
at around 26 T and the intensity increases with increasing magnetic field. 
On the other hand, the intensity of the reflection peak in the low-field phase (peak B) decreases with magnetic 
field above 26 T.
The field position where the intensity alternation between the peak A and B takes place is close to 
the metamagnetic transition field. \cite{Mush2}
This fact indicates that the observed peak shift to the higher diffraction angle (from B to A) corresponds to 
the lattice contraction due to the field-induced valence state transition. 
Moreover, it is found that the two peaks coexist in the vicinity of the transition field ; it strongly suggests that the observed 
transition is the first order transition.

\begin{figure}[t]
\begin{center}
\includegraphics[width=7cm]{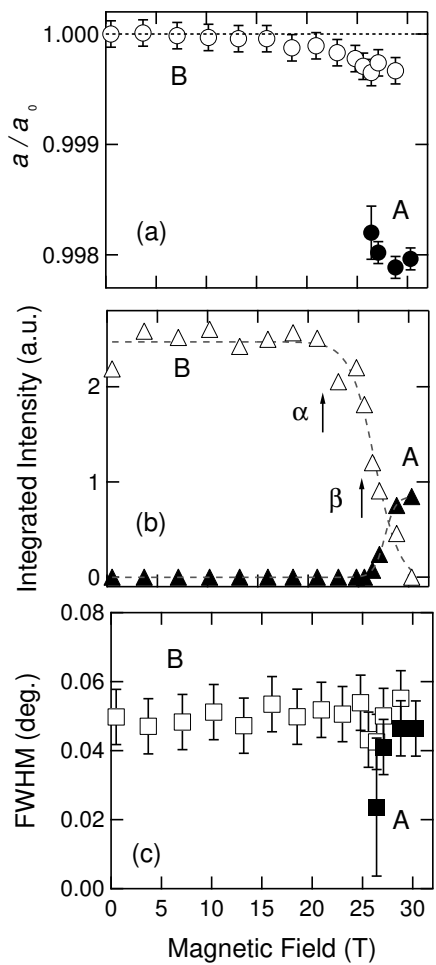}
\end{center}
\caption{The analyzed data at 32 K. (a) Relative change in the 
lattice constant deduced from the (220) reflection peak shift. 
(b) Integrated intensities (along the $\theta-2\theta$ direction) of the (220) reflection peak 
as a function of the magnetic field. The dashed curves are the guide for eyes. The two arrows labeled 
$\alpha$ and $\beta$ denote the onset and half-value positions of the metamagnetic transition, respectively.
The magnetization curve at 32 K is shown in Fig.\ref{magR}.
(c) Field dependence of the full width at the half maximum of the (220) reflection peaks.
The large error bar around the transition field corresponds to the large standard deviation value 
in the fitting analysis, which is due to the small intensity of the peak A.
}
\label{analysis}
\end{figure}

To examine the details of the lattice deformation due to the transition, 
we made a fitting analysis using the Gaussian curve and evaluated the peak positions, 
integrated (along 2$\theta$) intensities, and widths for the two peaks. (See Fig.~\ref{analysis}(a) $\sim$ (c).)
The resultant fitting curves are shown in Fig.~\ref{profile} by solid curves. 
Figure~\ref{analysis}(a) shows the field variation of the relative lattice constant 
deduced from the fitting. The change in the lattice constant by the transition
($(a_0 - a)/a_0 $) is found to be $-2.0 \pm 0.1 \times 10^{-3}$, 
where $a_0$ is the lattice constant at 32 K and zero magnetic field.

We estimated the lattice constant of the hypothetical zero-field localized (Yb$^{3+}$) phase at 32 K
using the linear extrapolation of the temperature dependence of the lattice constant into the 
low temperature region.
We used the temperature dependence reported previously. \cite{Lawrence}
The difference of the lattice constant 
$\Delta a$ between the itinerant (Yb$^{2.9+}$) phase and the hypothetical 
localized (Yb$^{3+}$) phase is evaluated to be $\Delta a/a_0=1.8 \pm0.1 \times 10^{-3}$ at 32 K. 
This is in good agreement with the value which we obtained from the field-induced lattice contraction. 
This fact suggests that the isothermal variation of the lattice constant 
at the valence state transition is observed in the present experiment.

It is interesting to point out that the lattice constant in the low-field-phase at 
near the transition field (20 $\sim$ 28 T)
is slightly 
smaller than that at far below the transition field. 
Similarly, the lattice constant of the high-field-phase at near the transition seems to be slightly 
larger than that above the transition field. Moreover, those lattice constants change continuously.
This experimental fact might be explained by a kind of environmental effect caused by the interactions between 
the domains of the different valence states, i.e., the lattice deforms when the two phases coexist and the 
degree of the deformation depends on the volume ratio of the two phases.

When we roughly estimate the magnetization energy gain by the magnetization jump at the 
metamagnetic transition using $\int {MdH} $ ($H$ is a magnetic field and $M$ denotes magnetization) 
the energy gain is found to be $1.5 \times 10^{6}$ J for unit volume (1 m$^3$)at 30 K. 
We used the magnetization jump $\sim 3 \mu_B$ for an Yb ion and the transition width was 
taken from 25 T to 30 T. \cite{Mush2}
Since the elastic energy loss
 $\frac{1}{2}\int {3(c_{11}+2c_{12})(\frac{\Delta a}{a})^2 dV} $ 
 ($c_{11}$ and  $c_{12}$ are the elastic constants; 17.2 $\times 10^{10}$ J/m$^3$ and 
8.29 $\times 10^{10}$ J/m$^3$ for $c_{11}$ and $c_{12}$, respectively.\cite{elastic}) 
for $\Delta a /a =1.8 \times 10^{-3}$ is $1.6 \times 10^{6}$ J 
the magnetiztion energy gain (corresponding to the Zeeman energy) and the elastic energy loss are 
found to be roughly balanced. 

\section{Discussions }
In the following, we discuss the interplay between the metamagnetism and the lattice contraction 
at the valence state transition.
In Fig.~\ref{analysis} (b) we compare the field dependence of the integrated intensity of the peak B 
with the magnetization at the same temperature.
It is found that the reduction of the reflection intensity and the increase of magnetization show 
one to one correspondence.
The similarity was also found in the down-sweep results (not shown in the figure).

In the simplest picture, the magnetization scales with the volume of the localized (Yb$^{3+}$) phase.
Although the integration was done only along 2$\theta$ it is also natural to assume the integrated 
intensity of the peak B represents the volume of the itinerant 
(Yb$^{2.8+ \sim 2.9+}$) phase, because the cubic symmetry is expected to be unchanged by the transition.
The observed correspondence between the magnetization and the Bragg reflection intensity supports this 
simplest picture.
In conventional first order phase transitions, such volume change is associated with the coexistence of 
corresponding two phases.
However, only peak B is found at the beginning of the metamagnetic transition and the peak A starts to grow 
in the \textit{middle} of the transition.
It seems that this behavior contradicts with the conventional first order phase transitions, i.e., the peak A should
appear at the field position where the intensity of the peak B starts to decrease.

\begin{figure}[t]
\begin{center}
\includegraphics[width=7cm]{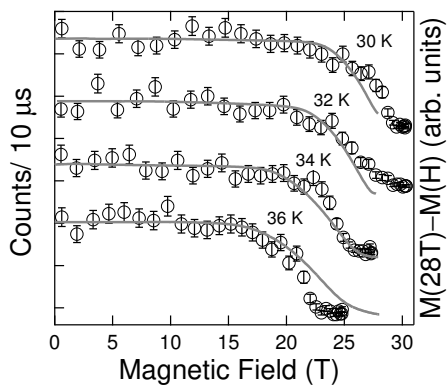}
\end{center}
\caption{Field dependence of the Bragg reflection peak intensities at different temperatures are shown 
for the up-sweep pulsed field (open circles).  
Scaled magnetization data $M$ are shown for the right vertical axis by gray solid curves, where $M$(28T) 
is the magnetization at 28 T for each temperature.
}
\label{magR}
\end{figure}

A possible interpretation of this phenomenon is that the domain size of the high-field-phase is very small 
at the beginning of the transition 
and it cannot be observed as a well-defined Bragg reflection peak.
It is found that
the peak width in the low-field-itinerant-phase $\Delta 2 \theta  \sim 0.05^{\circ}$ corresponds to 
the correlation length of the order of $10^3$ \AA $\;$.
If the size of the high-field-phase is much smaller than this length scale, it would 
be difficult to observe the Bragg reflection from such microscopic domains. 
The present observation may be the first experimental evidence of the microscopic nucleation in the valence 
state transition. Although the Bragg peak width seems to be unchanged as shown in Fig.\ref{analysis} (c) 
the mosaic structure of the crystal may cover the intrinsic peak widths which might be dependent on the magnetic field. 
In Fig.~\ref{magR} we plot the field dependence of the intensity of the peak B at different temperatures. 
The scaled magnetization curves are shown together. A relatively good agreement between the field dependencies 
of these two quantities supports the
above-mentioned picture, i.e., the valence transition starts to occur at the onset of the metamagnetic 
transition and correspondingly the intensity of the Bragg peak from the low-field-itinerant phase decreases.
%
%
%
%
\section{Summary}
We have performed high-magnetic-field x-ray diffraction experiments using a pulsed magnet and 
synchrotron x-rays for the first time. The structural phase transition due to the field-induced valence 
transition in YbInCu$_4$ has been studied by a direct observation of the lattice deformation. 
The coexistence of the two valence states 
was clearly observed at near the transition magnetic field ; 
it strongly suggests that the field induced valence transition is the first order transition. 

We discussed the interplay between the lattice contraction and the metamagnetism and proposed 
a microscopic nucleation picture at the beginning stage of the field-induced valence state transition.
Since a direct observation of the electronic states would shine some light on the validity of this model, 
high-field  x-ray absorption spectroscopy near the  $L$-edge of the Yb ions would be one of the most 
promising experiments.

\section{Acknowledgements}
This work was partly supported by the Ministry of Education, Science, Sports and Culture, 
Grant-in-Aid for Young Scientists (B),15740183 and the REIMEI Research Resources of 
Japan Atomic Energy Research Institute. The authors thank Y. Yoda and S. Kishimoto for valuable 
discussions on the time domain x-ray measurement technique using an avalanche photo diode.

\end{document}